\documentstyle[prb,aps,epsf,multicol,tabularx]{revtex}

\tighten
\begin{document}




\draft

\title{Conductance and density of states as 
the Kramers-Kronig dispersion relation}

\author{Tooru Taniguchi}

\address{D\'epartement de Physique Th\'eorique, Universit\'e 
de Gen\`eve, CH-1211, Gen\`eve 4, Switzerland}

\date{\today}

\maketitle

\begin{abstract}

   By applying the Kramers-Kronig dispersion relation to the 
transmission amplitude a direct connection of the conductance 
with the density of states is given in quantum scattering 
systems connected to two one-channel leads.  
   Using this method we show that in the Fano resonance the peak 
position of the density of states is generally different from the 
position of the corresponding conductance peak, whereas in the 
Breit-Wigner resonance those peak positions coincide. 
   The lineshapes of the density of states 
are well described by a Lorentz type in the both resonances. 
   These results are verified by another approach using a specific 
form of the scattering matrix to describe 
scattering resonances. 

\end{abstract}

\vspace{0.2cm}
\begin{multicols}{2}

\narrowtext

%
%
%
%

\newcommand{\scsc}{\scriptscriptstyle}
\newcommand{\BEQ}{\begin{eqnarray}}
\newcommand{\EEQ}[1]{\label{#1}\end{eqnarray}}
\newcommand{\caC}{{\cal C}}
\newcommand{\caD}{{\cal D}}
\newcommand{\caF}{{\cal F}}
\newcommand{\caK}{{\cal K}}
\newcommand{\caP}{{\cal P}}
\newcommand{\sign}{\mbox{sign}}

\makeatletter 
   \def\@biblabel#1{(#1)} 
   \def\refname{\vspace{0cm} 
   \noindent {\bf REFERENCES}  
   \vspace{0.5cm}} 
   \makeatother    

   
%
%
%
%
%
%
%

%
%
%

\section{Introduction}

   The developments of nano-scale fabrication technique 
made possible confining electrons in a small region so that 
the system shows a discrete energy spectrum. 
   Such an electron system is called the quantum dot, whose  
characteristics have been investigated in 
many theoretical and experimental works.\cite{kas93,kou97} 

   One way by which characteristics of a quantum 
dot can be investigated is to connect leads to it   
and to measure its conductance. 
   Many such experiments have actually been carried out and have 
shown sharp 
peaks of the conductance or the staircase structure in electric 
currents as a function of gate voltage or source-drain 
voltage.\cite{kas93,kou97,ree88,sco89,kou91,fox93}
   These experimental results about conductance peaks have been 
interpreted on the hypothesis that the electric current 
through the quantum dot occurs 
if there is at least one of the energy levels of the quantum dot between  
chemical potentials of the reservoirs connected to the quantum dot 
via leads. 
   This hypothesis is justified if the peak position of the 
conductance coincides with the corresponding peak position of the 
density of states in the quantum dot.  
   
   In this paper we investigate this hypothesis 
about the peak positions of the conductance and the density of states. 
   We consider a quantum dot connected to two one-channel leads, 
and assume that the system has a time-reversal symmetry. 
   In this system, from the scattering matrix the conductance 
and the density of states are calculated by using the Landauer 
conductance formula \cite{lan70,eco81,fis81,but85,but86} 
and the Friedel sum rule \cite{fri52,langar61,das69}, respectively. 
   Moreover, in order to connect the conductance with the density of 
states we use the Kramers-Kronig dispersion relation.   
   The Kramers-Kronig dispersion relation connects the real part 
of a function with its imaginary part by the Hilbert 
transformation, based on the analyticity of the function.  
   By applying this relation to the logarithm of a scattering matrix 
element we obtain formulas allowing us to calculate the conductance from 
the density of states and to calculate the density of states from 
the conductance.
   These formulas are used to investigate 
a relation of peak positions of the conductance and the 
density of states. 

   We consider two kinds of resonances which are 
called the Breit-Wigner resonance and the Fano resonance.   
   The Breit-Wigner resonance \cite{bre36} is characterized as the 
conductance lineshape 

\begin{figure}[t]
   \epsfxsize4.6cm  
   \vspace{-0.5cm}
   \centerline{\epsffile{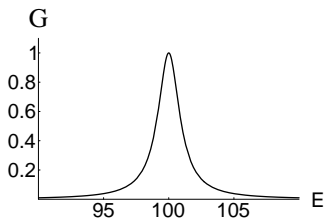}} 
   \vspace{-0.5cm}
   \caption{Conductance lineshape of the Breit-Wigner 
      resonance as a function of energy.}  
   \vspace{0cm}
   \label{ResonBreit} 
\end{figure}

\BEQ
   G_b(E) =  \Lambda_b \frac{1}{(E-E_{0})^2+\Delta^2}
\EEQ{BreitWigneLine}

\noindent of a Lorentz type around a resonant energy 
$E_{0}$ as a function of energy 
$E$, where $\Lambda_b$ is a positive constant. 
   Here the real constant $\Delta$ represents 
a coupling strength of the 
quantum dot with leads, and takes a small value compared  
with energy level spacings of the quantum dot 
in a weak coupling case with leads.  
   This resonance lineshape agreed  
with experimental results for conductance in some quantum 
dots.\cite{fox93} 
   Fig. \ref{ResonBreit} shows this conductance lineshape 
with the parameter values $E_{0}=100$, $\Delta=1$ and $\Lambda_b =1$.  
   On the other hand, the Fano resonance \cite{fan61} 
is characterized by the conductance lineshape

\begin{figure}[t]
   \epsfxsize4.6cm   
   \vspace{-0.5cm}
   \centerline{\epsffile{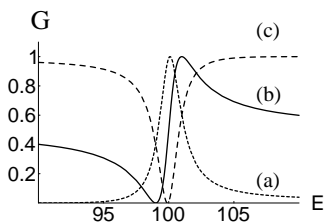}} 
   \vspace{-0.5cm}
   \caption{Conductance lineshapes of the Fano resonance 
      as functions of energy. 
      The graphs (a), (b) and (c)  
      are corresponding to the case of $Q=10$, $1$ and 
      $0.1$, respectively.}  
   \vspace{0cm} 
   \label{ResonFano} 
\end{figure}

\BEQ
   G_f(E) = \Lambda_f \frac{(E-E_{0}+Q)^2}{(E-E_{0})^2+\Delta^2}
\EEQ{FanoLine}

\noindent around a resonant energy $E_{0}$, where $\Lambda_f$ 
is a positive constant. 
   Here the parameter $Q$ determines asymmetry in the 
conductance lineshape of the Fano resonance. 
   The Fano resonance lineshapes are drawn in Fig. 
\ref{ResonFano} with the parameter values $E_{0}=100$ and $\Delta=1$.  
   Here, we chose the parameter $\Lambda_f$ 
as $\Delta^2/(\Delta^2+Q^2)$ so that the peak value 
of the conductance is one. 
   The Fano resonance is caused by coupling discrete states 
with continuous states, and exhibits conductance zero points 
like the energy point $E=E_{0}-Q$ in 
Eq. (\ref{FanoLine}).\cite{tek93,noc94} 
   It should be noted that the Fano resonance is attributed to 
the Breit-Wigner type in the case of $|Q/\Delta|>>1$ (See 
Fig. \ref{ResonFano} (a).). 
   The conductance lineshape of the Fano resonance is actually 
observed experimentally by using the scanning tunneling microscopy 
\cite{li98,ma98,man00} and in quantum dots \cite{goe99,zac00}. 
   These experimental results show that even the 
case of $|Q/\Delta|<<1$ like Fig. \ref{ResonFano} (c) can happen.    

    By applying our formula using the Kramers-Kronig dispersion 
relation to these two kinds of resonances we obtain the following 
results:  
   (1) In the Breit-Wigner resonance the peak position of the density 
of states coincides with the position $E=E_{0}$ of the conductance peak. 
   (2) In the Fano resonance the density of states are  
independent of value of the asymmetric parameter $Q$ in a 
weak coupling case with leads, and the peak position of the 
density of states is at $E=E_{0}$.   
   Eq. (\ref{FanoLine}) shows that the peak position of the 
conductance depends on the asymmetric parameter $Q$ and is 
given by $E=E_{0}+\Delta^2/Q$. 
   Therefore, in the case of $|Q/\Delta|>>1$ the peak position 
of the density of states is close to the position 
of the conductance peak. 
   On the other hand in the case of $|Q/\Delta|<<1$ 
the peak position of the density of states is rather 
close to the energy $E_0-Q$ at which the conductance takes a minimum value.
   We also show that in both resonance types 
the lineshapes of the density of states 
are a Lorentz type. 
   These results are correct even in the case where   
an electron-electron interaction like the charging effect inside 
the quantum dot plays an important role, 
because the Friedel sum rule is correct even in presence of   
electron-electron interactions.\cite{langar61}
   
   We verify the above results by another approach which 
does not use the Kramers-Kronig dispersion relation.  
   It is an approach using a specific form of the scattering 
matrix to describe scattering resonances. 
   We show that only the Breit-Wigner and the Fano resonances 
are derived from this scattering matrix. 
   By applying the Landauer conductance formula and the Friedel sum 
rule to this specific form of the scattering matrix we  
calculate the density of states and the conductance, and obtain 
the same results as with the dispersion relation approach.


\section{Kramers-Kronig Dispersion Relation in the Transmission 
Amplitude} 

   The system which we consider in this paper is the quantum dot 
connected to particle reservoirs via two one-dimensional leads. 
   We neglect the effect of a magnetic field 
so that the system has the time-reversal symmetry. 
   For such a system the scattering matrix $S(E)=(S_{ll'}(E))$ is 
represented as a $2\times 2$ symmetric and unitary matrix at 
any energy $E$.
   The conductance $G(E)$ is given by the Landauer conductance 
formula 

\BEQ
   G(E) = \frac{q^2}{2\pi\hbar} |t(E)|^2
\EEQ{Landa} 

\noindent with the charge $q$ of the particle,  
the Planck constant $2\pi\hbar$ and the 
transmission amplitude $t(E)\equiv S_{12}(E)(=S_{21}(E))$. 
    The density of states $\rho(E)$ is given by the Friedel sum rule 

\BEQ
   \rho(E)  = \frac{1}{\pi} 
   \frac{\partial \theta_{\!f} (E)}{\partial E}
\EEQ{Fried0}

\noindent which $\theta_{\!f}(E)$ is the Friedel phase defined by 

\BEQ
   \theta_{\!f}(E)  \equiv \frac{1}{2i} \ln \mbox{Det} \{S(E)\}.
\EEQ{FriedPhase}

\noindent It is important to note that the Friedel phase 
$\theta_{\!f}(E)$ and the transmission amplitude phase 
$\theta_{\!t} (E) (\equiv \mbox{Arg}\{t(E)\})$ are not completely 
independent. 
   Actually, if the conductance is not zero in any value of energy, 
then the transmission amplitude phase $\theta_{\!t}(E)$ is 
simply given by $\theta_{\!f} (E) + \pi/2$. 
   On the other hand, if the conductance takes zero in some  
energy points $E=E^{\scsc (n)}$, $n=1,2,\cdots$, then the 
transmission amplitude phase can have discontinuities of 
$\pm\pi$ in those points, and is connected to the Friedel 
phase $\theta_{\! f} (E)$ as $\theta_{\! t} (E) 
= \theta_{\! f} (E) + \eta(E)$ with $\eta(E) \equiv 
\nu + \pi \sum_{n} \gamma_n \Theta(E-E^{\scsc (n)})$.\cite{tan99}  
   Here $\nu$ is an energy-independent constant, 
$\Theta(x)$ is the step function of $x$, 
and $\gamma_n$ is a constant taking the value $-1$, 
$0$ or $1$ only.
   In this paper we treat the conductance like Eq. 
(\ref{FanoLine}), so we should make up our formula based on 
the case where there is a conductance zero point. 

   Since the transmission amplitude phase 
$\theta_{\! t} (E)$ has discontinuities in the conductance zero 
points, we should not assume that the logarithm of the transmission 
amplitude $t(E)$ itself is an analytic function of energy. 
   Therefore, in order to apply the Kramers-Kronig 
dispersion relation in which the analyticity of the function 
plays an essential role, we must carefully remove the singularity 
caused by the conductance zero points from the logarithm of the 
transmission amplitude $t(E)$. 
   For this purpose we represent the transmission amplitude $t(E)$ as 
   
 \BEQ
   t(E)=\lim_{\varepsilon\rightarrow +0} e^{i\eta(E)} 
   \cdot e^{2^{-1} \ln (\varepsilon+|t(E)|^2) +i \theta_{\! f}(E)}. 
\EEQ{TransAmpli} 

\noindent 
   The limit $\varepsilon\rightarrow +0$ is introduced to avoid the 
divergences of the function $\ln |t(E)|^2$ 
of $E$ in the conductance zero points.   
   In addition, the function $\theta_{\! f}(E)$ of $E$ is a continuous 
function because its derivative gives the density of states $\rho(E)$ 
multiplied by $\pi$, which should be a continuous function of energy.
   Therefore we get the function 
$2^{-1} \ln (\varepsilon+|t(E)|^2) +i \theta_{\! f}(E)$, which can be 
assumed to be a continuous function of energy.

   In the next step we separate its asymptotic form from the 
transmission amplitude and we make a function which goes to    
zero as the energy $E$ goes to infinity. 
   For this purpose we introduce the asymptotic forms of the 
functions $|t(E)|^2$ and $\theta_{\! f}(E)$ as 

\BEQ
   |t(E)|^2 \stackrel{E\rightarrow+\infty}{\sim} 
   T^{\scsc (\infty)}(E)
\EEQ{AsymTransProba}
\BEQ
   \theta_{\! f}(E) \stackrel{E\rightarrow+\infty}{\sim} 
   \theta_{\! f}^{\scsc (\infty)}(E)
\EEQ{AsymFriedPhase}

\noindent As an example of the asymptotic transmission amplitude,  
in the one-dimensional system we may take $t(E) 
\stackrel{E\rightarrow+\infty}{\sim} \exp(ikl)$, where $l$ is the 
length of the system and $k$ is the 
wave vector $\sqrt{2mE}/\hbar$ with the mass $m$ of the particle, 
so this gives $T^{\scsc (\infty)}(E)=1$ and 
$\theta_{\! f}^{\scsc (\infty)}(E)=kl$. 
  The transmission amplitude $t(E)$ is represented as

\BEQ
   t(E) &=& \lim_{\varepsilon\rightarrow +0} 
   e^{2^{-1} \ln (\varepsilon+T^{\scsc (\infty)}(E)) }
   \nonumber \\ 
   && \hspace{0.6cm} \times \; 
   e^{i \{\theta_{\! f}^{\scsc (\infty)}(E)+\eta(E)\}} \cdot 
   e^{\Phi_{\varepsilon}(E)}
\EEQ{Trans}

\noindent where $\Phi_{\varepsilon}(E)$ is the 
imaginary function defined by  

\BEQ 
   \Phi_{\varepsilon}(E) &\equiv& \frac{1}{2}
   \ln \frac{\varepsilon + |t(E)|^2}
   {\varepsilon + T^{\scsc (\infty)}(E)} 
   \nonumber \\ 
   && \hspace{0.6cm} 
   + i\left[ \theta_{\! f}(E)-\theta_{\! f}^{\scsc (\infty)}(E)\right]. 
\EEQ{Logar}

\noindent An important characteristic of the function 
$\Phi_{\varepsilon}(E)$ is that 
this function satisfies the condition  

\BEQ
  \lim_{E\rightarrow+\infty} \Phi_{\varepsilon}(E) = 0, 
\EEQ{Asymp2}

\noindent and can be assumed to be a continuous function of energy.  
  The real part of the function $\Phi_{\varepsilon}(E)$ 
gives the conductance 

\BEQ
   G(E)  =  
   \lim_{\varepsilon\rightarrow+0} G^{\scsc (\infty)}(E)
   e^{2 \mbox{Re}\{\Phi_{\varepsilon}(E)\}}
\EEQ{Landa1}

\noindent by using Eq. (\ref{Landa}), where $G^{\scsc (\infty)}(E)$ 
is the conductance 
$(q^2/(2\pi\hbar)) T^{\scsc (\infty)}(E)$ in the high 
energy limit. 
   Using Eq. (\ref{Fried0}) the density 
of states $\rho(E)$ is connected to  the imaginary part of
the function $\Phi_{\varepsilon}(E)$ by 

\BEQ
   \rho(E)  = \rho^{\scsc (\infty)}(E) + 
    \lim_{\varepsilon\rightarrow+0} \frac{1}{\pi} 
   \frac{\partial \mbox{Im}\{\Phi_{\varepsilon}(E)\}}{\partial E}
\EEQ{Fried1}

\noindent where $\rho^{\scsc (\infty)}(E)$ is the asymptotic form 
of the density of states in the high energy limit and is given by 
$ (1/\pi) \partial \theta_{\! f}^{\scsc (\infty)}(E)/\partial E$.
   Now we finish preparing the function 
$\Phi_{\varepsilon}(E)$ to which we apply 
the Kramers-Kronig dispersion relation.  

   So far, the function $\Phi_{\varepsilon}(E)$ has been 
defined only in the real energy region $(0,+\infty)$.  
   (Here we took the origin of energy so that the lower bound of 
the energy is zero.) 
   Now, in order to apply the Kramers-Kronig dispersion relation 
to the function $\Phi_{\varepsilon}(E)$, 
we extend this function  
so that it is defined in the whole upper half 
plain of the imaginary number $E$ including the real axis. 
   We assume that such an extension can be done under the three 
conditions:  

\begin{quote}
\begin{description}
\item[{\rm (I)}]  The function $\Phi_{\varepsilon}(E)$ 
   of $E$ is analytic in the whole upper half 
   plain and in the real axis in the imaginary number $E$.
\item[{\rm (I${\!}$I)}]   $\lim_{|E|\rightarrow+\infty} 
   |\Phi_{\varepsilon}(E)| = 0$ in any energy $E$ satisfying 
   $\mbox{Im} \{E\} \geq 0$.  
\item[{\rm(I${\!}$I${\!}$I)}] $\Phi_{\varepsilon}(-E) = 
   \Phi_{\varepsilon}(E)^*$ in 
   any real number $E$.
\end{description}
\end{quote}

\noindent It should be noted that the condition (I${\!}$I) is 
a generalization of Eq. (\ref{Asymp2}). 
   In this paper we choose the value $\theta_{\! f}(0) - 
\theta_{\! f}^{\scsc (\infty)}(0)$ as $0$, 
so that the right-hand side and the left hand side in the equation 
of the condition (I${\!}$I${\!}$I) coincide at the origin  
$E=0$.  
   Known as the Kramers-Kronig dispersion relation, 
\cite{joa75} using the conditions 
(I), (I${\!}$I) and (I${\!}$I${\!}$I) the real part and the imaginary part of 
the function $\Phi_{\varepsilon}(E)$ are connected as 

\BEQ
   \mbox{Re}\{\Phi_{\varepsilon}(E)\} = \frac{2}{\pi} \hat{\caP}\! 
   \int_0^{+\infty} dE' \; 
   \frac{E' \, \mbox{Im}\{\Phi_{\varepsilon}(E')\}}{E'^2-E^2} 
\EEQ{KrameKroni1} 
\BEQ
   \mbox{Im}\{\Phi_{\varepsilon}(E)\} = -\frac{2}{\pi} \hat{\caP}\! 
   \int_0^{+\infty} dE' \; 
   \frac{E \, \mbox{Re}\{\Phi_{\varepsilon}(E')\}}{E'^2-E^2}
\EEQ{KrameKroni2}   

\noindent where the operator $\hat{\caP}$ means to take the 
principal integral in the following integral. 
 
   Using Eqs.  (\ref{Logar}), (\ref{Landa1}) and (\ref{Fried1}), 
the relations (\ref{KrameKroni1}) and (\ref{KrameKroni2}) lead to 
a direct connection between the conductance and the density of states:     

\BEQ
   G(E) &=& G^{\scsc (\infty)}(E) 
   \exp \biggl\{-\int_{0}^{+\infty} dE' \; 
   \caC(E,E') 
   \nonumber \\ 
   && \hspace{2.4cm} \times \; \left[\rho(E') 
   - \rho^{\scsc (\infty)}(E')\right] \biggr\}
\EEQ{ConduDensi}
\BEQ
   \rho(E) &=& \rho^{\scsc (\infty)}(E) 
   +\lim_{\varepsilon\rightarrow+0}
   \int_{0}^{+\infty} dE' \; \caD(E,E') \nonumber \\ 
   && \hspace{3cm} \times \;\ln 
   \frac{\varepsilon + G(E')}{\varepsilon + G^{\scsc (\infty)}(E')}
\EEQ{DensiCondu}

\noindent where the functions $\caC (x,y)$ and $\caD (x,y)$ of $x$ 
and $y$ are defined by 

\BEQ
   && \caC(x,y) 
   \nonumber \\ 
   && \hspace{0cm}  \equiv 
   \lim_{\epsilon\rightarrow+0}  
   \ln \left\{\left[ (x-y)^2 +\epsilon^2 \right] 
   \left[(x+y)^2 + \epsilon^2 \right] \right\}
\EEQ{FunctC}
\BEQ
   && \caD(x,y) 
   \equiv - \lim_{\epsilon\rightarrow 0} 
   \frac{1}{2\pi^2}  \left\{
   \frac{(x-y)^2-\epsilon^2}
   {\left[(x-y)^2+\epsilon^2\right]^2}\right.
   \nonumber \\ 
   && \hspace{3.6cm} + \left. \frac{(x+y)^2-\epsilon^2}
   {\left[(x+y)^2+\epsilon^2\right]^2}
   \right\}. 
\EEQ{FunctD}

\noindent (See Appendix A about the derivations of these equations.)
  Here, in order to derive Eq. (\ref{ConduDensi}) we assumed 
$\lim_{E\rightarrow +\infty} 
\{\theta_{\! f}(E) - \theta_{\! f}^{\scsc (\infty)}(E)\}\ln E = 0$, 
which is stronger than the condition (\ref{AsymFriedPhase}). 
   Eqs. (\ref{ConduDensi}) and (\ref{DensiCondu}) are the key 
results of this paper. 

   As a general feature of the conductance shown by using Eq. 
(\ref{ConduDensi}) the conductance $G(E)$ is invariant under the 
change $\rho(E) \rightarrow \rho(E) + 
\alpha$ (So $\rho^{\scsc (\infty)}(E) \rightarrow 
\rho^{\scsc (\infty)}(E) + \alpha$.) of the density of states 
in any constant $\alpha$.
   Similarly Eq. (\ref{DensiCondu}) implies 
that the density of states $\rho(E)$ is invariant under the change 
$G(E)\rightarrow \beta \; G(E)$ in any constant 
$\beta$.


\section{Application to the Breit-Wigner and Fano Resonance}

   In this section, using Eq. (\ref{DensiCondu}) we calculate the 
densities of states in the Breit-Wigner resonance 
and the Fano resonance.  
   In the actual calculation we use the equation 

\BEQ
   && \rho(E)  - \rho^{\scsc (\infty)}(E) 
      \nonumber \\ 
   && \hspace{0.6cm} =  - \lim_{\varepsilon\rightarrow+0} 
      \frac{1}{\pi^2}
      \hat{\caP}\!\int_0^{+\infty} \! dE' \; 
      \frac{E'}{E'^2-E^2} 
      \nonumber \\ 
   && \hspace{4cm}  \times \; 
      \frac{\partial}{\partial E'}
      \ln  \frac{G_{\varepsilon}(E')}{G_{\varepsilon}^{\scsc (\infty)}(E') }
\EEQ{DensiCondu2}

\noindent with $G_{\varepsilon}(E) \equiv \varepsilon + G(E)$ and 
$G_{\varepsilon}^{\scsc (\infty)}(E) \equiv 
\varepsilon + G^{\scsc (\infty)}(E)$. 
   Eq. (\ref{DensiCondu2}) is equivalent with Eq. (\ref{DensiCondu}), 
as shown in the end of Appendix A. 
 
   Before calculating the density of states, we consider 
some problems in applications of the formula 
(\ref{DensiCondu2}) to the conductances (\ref{BreitWigneLine}) 
and (\ref{FanoLine}). 
   First, strictly speaking, in order to obtain the density 
of states using the formula (\ref{DensiCondu2}) we need to 
know the value of the conductance in any energy $E$. 
   On the other hand Eqs. (\ref{BreitWigneLine}) and  
(\ref{FanoLine}) are correct only around the resonant  
energy $E_{0}$. 
   However the integral kernel $E'/(E'^2-E^2)$ in the formula 
(\ref{DensiCondu2}) has a large absolute value only around $E'=E$, 
so the value of conductance around the energy $E_{0}$ is enough to 
obtain approximately
the density of states around the energy $E_{0}$. 

   The second problem in applications of the formula 
(\ref{DensiCondu2}) is that we do not know the general 
asymptotic forms of the conductance and the density of states, 
which is needed to calculate the exact form of the density of states 
$\rho (E)$ by using Eq. (\ref{DensiCondu2}). 
   In this section we assume that the energy dependence of the 
asymptotic form of the transmission amplitude 
is the same as the one-dimensional case, namely $t(E) 
\stackrel{E\rightarrow+\infty}{\sim} 
\exp(i\lambda\sqrt{E})$ using a constant $\lambda$. 
   Therefore the asymptotic form of the conductance and the 
density of states are given by 
$G^{\scsc (\infty)}(E)=q^2/(2\pi\hbar)$ and 
$\rho^{\scsc (\infty)}(E) = \lambda/(2\pi\sqrt{E})$, respectively.   
   We do not have to care whether the conductances  
(\ref{BreitWigneLine}) and (\ref{FanoLine}) satisfy 
the condition $\lim_{E\rightarrow\infty} G(E) = q^2/(2\pi\hbar)$, 
because these forms of the conductances are justified only 
around the resonant energy $E_{0}$.  

   It is valuable to extract 
an essential part giving a peak of the density of states 
from the right-hand side of Eq.  (\ref{DensiCondu2}). 
   For this purpose we rewrite Eq. (\ref{DensiCondu2}) as 
   
\BEQ
   && \rho(E)   
      = - \lim_{\varepsilon\rightarrow+0} \frac{1}{2\pi^2}
      \hat{\caP}\!\int_{-\infty}^{+\infty} \! dE'  \;
      \frac{1}{E'-E} 
      \frac{\partial \ln  G_{\varepsilon}(E')}{\partial E'} \nonumber \\ 
   && \hspace{2cm} + \caF(E) + \frac{\lambda}{2\pi\sqrt{E}}
\EEQ{DensiCondu3} 

\noindent Here we used the specified asymptotic form 
of the conductance and the density of states, and 
$\caF(E)$  is defined by 

\BEQ
   && \caF(E) \equiv - \lim_{\varepsilon\rightarrow+0} 
      \frac{1}{2\pi^2} \int_{0}^{+\infty} \! dE'\;  
      \frac{\Xi_{\varepsilon}(E')}{E'+E}    
\EEQ{RemaiFunct}

\noindent with $\Xi_{\varepsilon}(E)\equiv 
(\partial/\partial E) \ln [G_{\varepsilon}(E) / G_{\varepsilon}(-E)]$. 
   The function $\caF(E)$ of $E$ is estimated as 
   
\BEQ
   && \left|\caF(E)\right| < \lim_{\varepsilon\rightarrow+0} 
      \frac{1}{2\pi^2} \frac{1}{E} 
       \int_{0}^{+\infty} \! dE'\; \left| \Xi_{\varepsilon}(E') 
      \right|.  
\EEQ{RemaiFunct2}  

\noindent We consider a weak coupling case of the quantum dot 
with leads, so we may regard the constant $\Delta$ as a 
small parameter compared  
with energy level spacings of the quantum dot.
   In this case we can assume that the energy value $E_{0}$ 
is large enough compared with the constant $|\Delta|$.  
   Noting that it is enough for us to calculate the 
density of states $\rho(E)$ only around the energy $E_{0}$, 
we estimate that the contribution 
of the function $\caF(E)$ to the density of states 
is negligible around the energy $E_{0}$
under the condition that the integral $\int_{0}^{+\infty} \! dE \;
|\Xi_{\varepsilon}(E)|$ has a finite value, 
because of the small factor 
$1/E\simeq 1/E_{0}$ in the right-hand side 
of Eq. (\ref{RemaiFunct2}).    
   The third term in the right-hand side of Eq. (\ref{DensiCondu3}) 
is a monotonous decreasing function of energy, so this part is 
also negligible in a large energy value $E \simeq E_{0}$ and 
almost does not 
contribute to changes of the peak position and the configuration of 
the density of states.    
   Therefore the main contribution to the peak 
of the density of states comes only from the first term in 
the right-hand side of Eq. (\ref{DensiCondu3}). 
   

\subsection{Breit-Wigner resonance}
  
\begin{figure}[t]
   \epsfxsize4.6cm
   \vspace{-0.5cm}
   \centerline{\epsffile{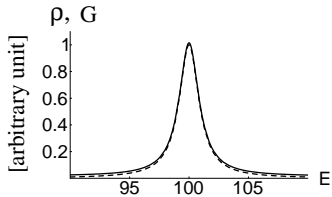}} 
   \vspace{-0.5cm}
   \caption{Density of states (solid line) and the conductance 
   (dashed line) as functions of energy in the Breit-Wigner 
   resonance.} 
   \vspace{0cm} 
 \label{DensiBreit} 
\end{figure}

   Fig. \ref{DensiBreit} is the Breit-Wigner resonance lineshape 
(\ref{BreitWigneLine}) and the corresponding density of states 
which is calculated by using Eq. (\ref{DensiCondu2}). 
   Here we chose the parameters as  
$\lambda =1$, and the other parameter values are the same as in   
Fig. \ref{ResonBreit}.
   Fig. \ref{DensiBreit} shows that the peak position of the 
density of states coincides with the peak position of the 
conductance in the Breit-Wigner resonance. 

   Now we check this result by the analytical 
consideration based on Eq. (\ref{DensiCondu3}) neglecting 
its second and third terms. 
   Substituting Eq. (\ref{BreitWigneLine}) into 
Eq. (\ref{DensiCondu3}) we obtain the density of states as 

\BEQ 
   \rho(E) &\simeq& \frac{1}{\pi^2}
      \hat{\caP}\!\int_{-\infty}^{+\infty} \! dE'  \;
      \frac{1}{E'-E} 
      \frac{E'-E_{0}}{(E'-E_{0})^2+\Delta^2} \nonumber \\ 
   &=& \frac{1}{\pi} 
      \frac{|\Delta|}{(E-E_{0})^2+\Delta^2}. 
\EEQ{DensiBreitAnali}
 
\noindent This implies that the density of states is a Lorentz type 
whose peak position is at $E=E_{0}$ and is independent of the 
value of the prefactor $\Lambda_b$ in the conductance lineshape 
(\ref{BreitWigneLine}).

%
\subsection{Fano resonance}

   The conductance (\ref{FanoLine}) in the Fano resonance is an 
example in which a conductance zero occurs, 
so the infinitesimal constant $\varepsilon$ in 
Eq. (\ref{DensiCondu2}) plays an important 
role in calculating the density of states.  
 
\begin{figure}[t]
   \epsfxsize4.6cm 
   \vspace{-0.5cm}
   \centerline{\epsffile{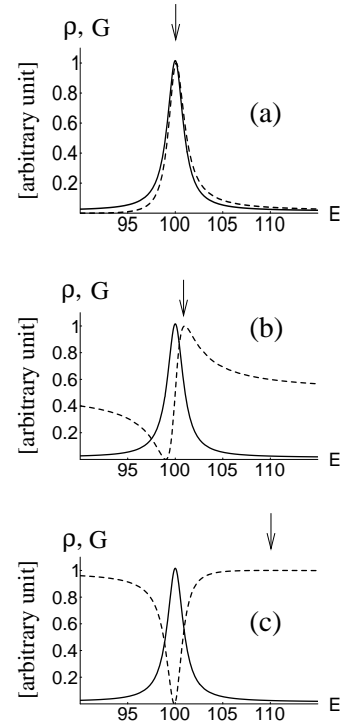}} 
   \vspace{-0.5cm}
   \caption{Density of states (solid line) and the conductance 
     (dashed line) as functions of energy in the Fano resonance. 
      The graphs (a), (b) and (c) are corresponding to the cases 
      of $Q=10$, $1$ and $0.1$, respectively. 
      The arrow in each graph shows the position of the 
      conductance peak.}
   \vspace{0cm} 
 \label{DensiFano} 
\end{figure}

   Fig. \ref{DensiFano} is the density of states 
corresponding to the Fano resonance lineshape (\ref{FanoLine}), 
which is calculated by using Eq. (\ref{DensiCondu2}).\cite{note}  
   Here, we chose the parameters as 
$\lambda =1$ and the other other 
parameter values are the same as in Fig. \ref{ResonFano}. 
   In the case of $|Q/\Delta|>>1$ (See 
Fig. \ref{DensiFano} (a).), where the 
conductance lineshape is close to the Breit-Wigner type,  
the peak position of the density of states is close to 
the peak position of the conductance. 
   On the other hand, in the case of $|Q/\Delta|<<1$ (See 
Fig. \ref{DensiFano} (c).), 
where the conductance lineshape is a gulf rather than a peak,    
the peak position of the density of states is rather close to 
the gulf of the conductance.   

  Now we calculate the density of states $\rho(E)$ by using Eq.     
(\ref{DensiCondu3}) neglecting its second and third terms. 
   After a small calculation (See Appendix B for the detail of 
the calculation) we obtain $\rho(E) \simeq 
(1/\pi)|\Delta|/\{(E-E_{0})^2+\Delta^2\}$, which is a Lorentz type 
with the peak at the energy $E_{0}$ and is the same as with the 
case of the Breit-Wigner resonance. 
   It should be emphasized that this form of the density of states 
is independent of the value of the asymmetric parameter $Q$ and 
the prefactor $\Lambda_f$ in the conductance lineshape (\ref{FanoLine}).  

\section{Another Approach using a Specific Form of Scattering Matrix} 

   In this section, using another approach which does not use the 
Kramers-Kronig dispersion relation, we verify our results 
 obtained in the previous section. 

   The specific form of the scattering matrix 

\BEQ
   S(E) =  A \left(I + i B \frac{1}{E-E_{0} +i\Delta} \right)  
\EEQ{SpeciScatt} 

\noindent around a resonant energy 
$E_{0}$ has been proposed to describe scattering  
resonances.\cite{jay72,lee00} 
   Here, the $2\times 2$ matrix $A \equiv (A_{ll'})$ is 
introduced as an energy-independent scattering matrix in the high 
energy limit, so that it is an unitary matrix in itself: 
$AA^\dagger=A^\dagger A = I$. 
   The matrix $B \equiv (B_{ll'})$ is also an energy-independent 
$2 \times 2$ matrix and satisfies the conditions 

\BEQ
   B^\dagger=B \;\mbox{and}\; B(B +2 I \Delta) = 0, 
\EEQ{CondiB}

\noindent so that the scattering matrix $S(E)$ given by Eq. 
(\ref{SpeciScatt}) becomes an unitary matrix in any energy $E$.  
   It may be noted that either $B = -2 I \Delta$ or $B=0$ satisfy the 
condition 
(\ref{CondiB}), but this gives an energy-independent conductance  
which is not pertinent to the subject of this paper. 
  Therefore in this section we assume $B \neq -2 I \Delta$ and $B\neq 0$, 
which lead to the parameterized representation 

\BEQ 
   B = \Delta \cdot 
   \left( \begin{array}{cc}
   -1+\sin\phi & e^{i\varphi}\cos\phi \\
   e^{-i\varphi}\cos\phi & -1-\sin\phi
   \end{array}\right) 
\EEQ{CondiB'}

\noindent  of the matrix $B$ with real parameters $\phi$ and 
$\varphi$.

   By applying the Friedel sum rule (\ref{Fried0}) to the 
scattering matrix (\ref{SpeciScatt}) the density of 
states $\rho(E)$ is given by 

\BEQ
    \rho(E) = \frac{1}{\pi} 
    \frac{\Delta}{(E-E_{0})^2+\Delta^2} , 
\EEQ{DensiState}

\noindent which satisfies  
$\lim_{\Delta\rightarrow+0} \rho(E)=\delta(E-E_{0})$.\cite{note2}
   Therefore the resonance lineshape of 
the density of states is always a Lorentz type with 
a peak at the energy $E_{0}$. 
   This agrees with the result concerning the lineshape of the 
density of states in the previous section.   

   Now we check that the scattering matrix (\ref{SpeciScatt}) 
gives the conductance lineshapes of the Breit-Wigner 
and the Fano resonance, and consider a relation between the peak 
positions of the conductance and the density of states. 
   First we consider the case of $A_{12}=0$. 
   In this case using Eqs. (\ref{Landa}) and 
(\ref{SpeciScatt}) the conductance 
$G(E)$ is represented as the Breit-Wigner type (\ref{BreitWigneLine})  
with $\Lambda_b=q^2|B_{12}|^2/(2\pi\hbar)$ by noting $|A_{11}|=1$, 
and the density of states is connected to the conductance simply as $
\rho(E)=\Delta \cdot G_b(E)/(\pi\Lambda_b)$.
   In this case the peak position of 
the conductance is at $E=E_{0}$ and coincides with 
the peak position  
of the density of states.   

   Second we consider the case of $A_{12}\neq 0$. 
   In this case, by applying the Landauer conductance formula 
(\ref{Landa}) to the scattering matrix (\ref{SpeciScatt}) we obtain  
the conductance 

\BEQ
   G(E) = W_f + \Lambda_f\frac{(E-E_{0}+Q)^2}{(E-E_{0})^2 + \Delta^2}
\EEQ{FanoLineSpeci}

\noindent of the Fano type with parameter values 

\BEQ
    W_f = \frac{q^2}{2\pi\hbar} |A_{12}|^2 \caK,
\EEQ{ConduFano1} 
\BEQ  
    \Lambda_f =  \frac{q^2}{2\pi\hbar} |A_{12}|^2 (1-\caK),  
\EEQ{ConduFano2} 
\BEQ  
    Q  = - \frac{d_1}{1-\caK}.  
\EEQ{ConduFano3}  

\noindent Here $\caK$ is defined by 

\BEQ
   \caK \equiv \frac{\Delta^2+d_1^2+d_2^2  
   -\sqrt{(\Delta^2+d_1^2-d_2^2)^2+4 d_1^2 d_2^2}}{2\Delta^2} 
\EEQ{Cak}

\noindent and $d_{j}, j=1,2$ are introduced as 

\BEQ
    d_1 \equiv \mbox{Im} \left\{  
    \frac{A_{11}}{A_{12}} B_{12} \right\}, 
\EEQ{d1} 
\BEQ   
    d_2 \equiv \Delta + B_{22}+
    \mbox{Re}\left\{\frac{A_{11}}{A_{12}} B_{12} \right\}.
\EEQ{d2}
 
\noindent  It is important to note that the constant $\caK$ 
satisfies 
the inequality $0\leq\caK\leq 1$ so the constants $W_f$ and 
$\Lambda_f$ given by Eqs. (\ref{ConduFano1}) and (\ref{ConduFano2}) 
are not negative.  
   The peak position of the conductance in this case is at 
$E=E_{0}+\Delta^2/Q$, 
which does not coincide with the peak position $E=E_{0}$ 
of the density of states shown by Eq. (\ref{DensiState}). 
   The Fano conductance (\ref{FanoLineSpeci}) takes a minimum 
value at the energy $E_{0}-Q$. 
   Therefore the peak position $E=E_{0}$ of the density of states 
is close to the peak position of the conductance 
in the case of $|Q/\Delta|>>1$, but as the 
quantity $|Q/\Delta|$ goes to $0$ it moves closer to the 
energy at which the conductance takes a minimum value. 
   This is exactly the same result as in the previous section.   

   The above results in this section are 
independent of the time-reversal symmetry of the system and are 
correct even in presence of a magnetic field. 
   However if the system has the 
time reversal symmetry and the conditions $S_{12}=S_{21}$ and 
$A_{12}=A_{21}$ are satisfied, then we obtain  $W_f=0$, $\Lambda_f 
=  (q^2/(2\pi\hbar)) |A_{12}|^2$ and  $Q  = - d_1$ 
   (See Appendix C for their proofs.). 
   Therefore the conductance (\ref{FanoLineSpeci}) becomes exactly 
the same form as Eq. (\ref{FanoLine}) in the time-reversal 
symmetric system.


\section{Conclusion and Remarks}

   In this paper using the Kramers-Kronig dispersion relation, 
the Landauer conductance formula and the Friedel sum rule 
we have discussed a method to calculate the density of states 
from conductance and to calculate  
conductance from the density of states in quantum scattering 
systems connected to two one-channel leads. 
   We considered the case of no magnetic field, so that the system 
had the time-reversal symmetry. 
   Our formula was applied to the Breit-Wigner resonance 
and the Fano resonance, and led to their profiles of 
the density of states. 
   In the Breit-Wigner resonance the peak positions of 
the conductance and the density of states coincide.   
   On the other hand, in the Fano resonance,  a relation of the 
peak positions of the conductance and the density of states depends 
on the parameter $|Q/\Delta|$ which determines asymmetry of 
the conductance lineshape. 
   In the case of $|Q/\Delta|>>1$ the peak position of the density of 
states is close to the position of the conductance peak, 
like the Breit-Wigner resonance. 
   However in the case of $|Q/\Delta|<<1$ the peak position of the 
density of states is rather close to the energy at which  
the conductance takes a minimum value. 
   We also showed that the lineshape of the density of 
states is a Lorentz type in both resonances. 
   These results are model-independent, and are correct even if 
electron-electron interaction inside the quantum dot 
plays an important role. 
   These results were verified by another consideration 
which does not use the Kramers-Kronig dispersion relation but 
uses a specified form of the scattering matrix to describe  
scattering resonances. 

   The relation between peak positions of the conductance 
and the density of states is important to explain 
an in-phase characteristic of the transmission amplitude 
phase (See Eq. (13) in Ref. \onlinecite{tan99}), 
which has been measured actually in an experiment.\cite{sch97} 
   Some works indicated that the Fano resonance property is 
important to cause this phenomenon.\cite{xu98,ryu98} 
   In Ref. \onlinecite{tan99} it has already been shown 
that in a simple model consisting of a branch connected 
to a one-dimensional perfect wire the peak positions of the density 
of states are in gulfs of  the conductance.   
 
   The advantage of our approach using the Kramers-Kronig dispersion 
relation is that we can know the density of states directly from 
the conductance which can be measured in experiments. 
   We can also calculate the density of states from the scattering 
matrix itself, but it is extremely difficult for the scattering matrix 
itself to be measured in experiments. 
   On the other hand, one of the disadvantages of the dispersion relation 
approach is that this approach is justified only under  
some restrictive conditions, for example, two one-channel leads, 
no magnetic field, the conditions (I), (I${\!}$I) and (I${\!}$I${\!}$I), etc. 
   We would get a wrong result if we neglected these conditions. 
   For example, the approach in the section 4 predicts a nonzero 
constant $W_f$ in presence of a magnetic field, and 
if we were to apply the formula (\ref{DensiCondu}) to such a nonzero 
$W_f$ case then the density of states would take a negative 
value in an energy region, which is not correct.
   To reduce the number of 
these conditions in our dispersion relation 
approach is one of the important future problems. 



\acknowledgments

%

   I wish to thank M. B\"uttiker for providing  
a stimulating environment for the present work. 
   I acknowledge a careful reading of this paper by M. Honderich.



\setcounter{section}{0} 
\makeatletter 
   \@addtoreset{equation}{section} 
   \makeatother 
   \def\theequation{\Alph{section}.%
   \arabic{equation}} 
 
\appendix
\section{Derivation of the connection between the conductance and 
the density of states}  

In this appendix we give the derivation of Eqs. 
(\ref{ConduDensi}), (\ref{DensiCondu}) and (\ref{DensiCondu2}). 

   First we should notice the equation 

\BEQ 
   && \hat{\caP}\frac{1}{E'^2-E^2}
      \nonumber \\
   && \hspace{0.3cm}= \frac{1}{2E'} \left(\hat{\caP}\frac{1}{E'-E} 
      +\hat{\caP}\frac{1}{E'+E} \right)  
      \nonumber \\
   && \hspace{0.3cm} = \lim_{\epsilon\rightarrow 0} 
      \frac{1}{2E'} \left[ \frac{E'-E}{(E'-E)^2+\epsilon^2}\right.
      \nonumber \\
   && \hspace{3cm} 
      + \left. \frac{E'+E}{(E'+E)^2+\epsilon^2} \right]. 
\EEQ{Princ} 

\noindent Similarly we obtain

\BEQ 
   &&\hat{\caP}\frac{1}{E'^2-E^2}
   \nonumber \\ 
   && \hspace{0.3cm} = \lim_{\epsilon\rightarrow 0} 
   \frac{1}{2E} \left[ \frac{E'-E}{(E'-E)^2+\epsilon^2} \right. 
   \nonumber \\
   && \hspace{3cm} 
      \left.- \frac{E'+E}{(E'+E)^2+\epsilon^2} \right].
\EEQ{Princ2}     

\noindent It follows from  Eqs. (\ref{Logar}), 
 (\ref{Landa1}),  (\ref{KrameKroni1}), 
(\ref{FunctC}) and (\ref{Princ}) that  

\BEQ
   && \ln\frac{G(E)}{G^{\scsc (\infty)}(E)} =  
      \lim_{\varepsilon\rightarrow+0} 
      2 \mbox{Re}\{\Phi_{\varepsilon}(E)\}  \nonumber \\
   && \hspace{0.3cm} = \lim_{\varepsilon\rightarrow+0} 
      \frac{4}{\pi} \hat{\caP}\! \int_0^{+\infty} dE' \; 
      \frac{E' \, \mbox{Im}\{\Phi_{\varepsilon}(E')\}}{E'^2-E^2} 
      \nonumber \\
   && \hspace{0.3cm}= \lim_{\varepsilon\rightarrow+0} 
      \lim_{\epsilon\rightarrow 0} 
      \frac{2}{\pi} \int_0^{+\infty} dE' \;
      \left[ \frac{E'-E}{(E'-E)^2+\epsilon^2} \right.  
      \nonumber \\
   && \hspace{2cm} + \left. \frac{E'+E}{(E'+E)^2+\epsilon^2} 
      \right]  \mbox{Im}\{\Phi_{\varepsilon}(E')\} 
      \nonumber \\
   && \hspace{0.3cm}=  \frac{1}{\pi}\int_0^{+\infty} dE' \;
      \frac{\partial \caC(E,E')}{\partial E'} 
      [\theta_{\! f}(E')-\theta_{\! f}^{\scsc (\infty)}(E')]
       \nonumber \\
   && \hspace{0.3cm}=  -\int_0^{+\infty} dE' \;\caC(E,E')
      \frac{1}{\pi}\frac{\partial 
      [\theta_{\! f}(E')-\theta_{\! f}^{\scsc (\infty)}(E')]}
      {\partial E'} 
      \nonumber \\
   && \hspace{0.3cm}= -\int_0^{+\infty} dE' \;\caC(E,E')
      \left[\rho(E)  - \rho^{\scsc (\infty)}(E)\right]
\EEQ{Landa2} 

\noindent where we used the conditions 
$\theta_{\! f}(0)-\theta_{\! f}^{\scsc (\infty)}(0)=0$ and  
$\lim_{E'\rightarrow +\infty} \caC(E,E')\, 
[\theta_{\! f}(E')-\theta_{\! f}^{\scsc (\infty)}(E')]=0$. 
   This leads to Eq. (\ref{ConduDensi}).
   Similarly, using Eqs. (\ref{Logar}), 
(\ref{Fried1}), (\ref{KrameKroni2}), (\ref{FunctD}) 
and (\ref{Princ2}) we obtain

\BEQ
   && \rho(E)  - \rho^{\scsc (\infty)}(E)  = 
      \lim_{\varepsilon\rightarrow+0} \frac{1}{\pi} 
      \frac{\partial 
      \mbox{Im}\{\Phi_{\varepsilon}(E)\}}{\partial E} 
      \nonumber\\
   && \hspace{0.3cm} = - \lim_{\varepsilon\rightarrow+0} 
      \frac{2}{\pi^2} \frac{\partial }{\partial E} \hat{\caP}\! 
      \int_0^{+\infty} dE' 
      \; \frac{E \, 
      \mbox{Re}\{\Phi_{\varepsilon}(E')\}}{E'^2-E^2}
      \nonumber \\
   && \hspace{0.3cm} = - \lim_{\varepsilon\rightarrow+0} 
      \lim_{\epsilon\rightarrow 0} 
      \frac{1}{\pi^2} \frac{\partial }{\partial E} 
      \int_0^{+\infty} dE' \;\left[ 
      \frac{E'-E}{(E'-E)^2+\epsilon^2} \right. 
      \nonumber \\
   && \hspace{2cm} \left. - \frac{E'+E}{(E'+E)^2+\epsilon^2} 
      \right] 
      \mbox{Re}\{\Phi_{\varepsilon}(E')\}  \label{Mid} 
      \\
   && \hspace{0.3cm} = \lim_{\varepsilon\rightarrow+0} 
      \int_0^{+\infty} dE'\; \caD(E,E') 
      \ln\frac{\varepsilon + G(E')}{\varepsilon 
      + G^{\scsc (\infty)}(E')}.
\EEQ{Fried2}
 
\noindent This leads to Eq. (\ref{DensiCondu}). 
   
   Using the expression (\ref{Mid}) we obtain 
another expression for a relation between the density of states and 
the conductance:

\BEQ
   && \rho(E) - \rho^{\scsc (\infty)}(E) 
      \nonumber \\ 
   && \hspace{0.3cm} = \lim_{\varepsilon\rightarrow+0} 
      \lim_{\epsilon\rightarrow 0} 
      \frac{1}{\pi^2} 
      \int_0^{+\infty} dE' \;  
      \nonumber \\ 
   && \hspace{2cm} \times \;  \left\{\frac{\partial }{\partial E'} 
      \left[ 
      \frac{E'-E}{(E'-E)^2+\epsilon^2} \right.\right. 
      \nonumber \\
   && \hspace{2.5cm} \left.\left. 
      + \frac{E'+E}{(E'+E)^2+\epsilon^2}  \right]\right\} \
      \mbox{Re}\{\Phi_{\varepsilon}(E')\} \nonumber \\ 
   && \hspace{0.3cm} =  \lim_{\varepsilon\rightarrow+0} 
      \int_0^{+\infty} dE' \; \tilde{\caD}(E,E') 
      \nonumber \\ 
   && \hspace{3cm} \times \; 
      \frac{\partial}{\partial E'} \ln 
      \frac{\varepsilon + G(E')}
      {\varepsilon + G^{\scsc (\infty)}(E')} 
\EEQ{Fried3}

\noindent where $\tilde{\caD}(E,E')$ is defined by 

\BEQ
   \tilde{\caD}(E,E') &\equiv& - 
   \lim_{\epsilon\rightarrow 0} 
   \frac{1}{2\pi^2}  \left[ \frac{E'-E}{(E'-E)^2+\epsilon^2} 
    \right. \nonumber \\ 
   && \hspace{2cm}  \left.
   + \frac{E'+E}{(E'+E)^2+\epsilon^2} \right] 
\EEQ{FunctTildD}

\noindent and satisfies the condition $\tilde{\caD}(E,0)=0$.
   Eqs. (\ref{Princ}) and (\ref{Fried3}) lead 
to Eq. (\ref{DensiCondu2}).


\section{Density of states in the Fano resonance}  

In this appendix we calculate the density of states by 
using Eqs. (\ref{FanoLine}) and (\ref{DensiCondu3}).
   Neglecting its second and third term, Eq. (\ref{DensiCondu3}) 
leads to the density of states $\rho(E)$:   
     
\BEQ 
   \rho(E)   
   \simeq - \lim_{\varepsilon\rightarrow+0} 
   \frac{1}{2\pi^2}
   \hat{\caP}\!\int_{-\infty}^{+\infty} \! dE'  \;
      \frac{ \Gamma_{\varepsilon}(E')}{E'+E_{0}-E}  
\EEQ{DensiConduFano2} 

\noindent where the function $\Gamma_{\varepsilon}(E)$ of $E$ is introduced as 

\BEQ
   \Gamma_{\varepsilon}(E) &\equiv&  
      \frac{1}{G_{\varepsilon\Lambda_f}(E+E_{0})} 
      \frac{\partial G_{\varepsilon\Lambda_f}(E+E_{0})}{\partial E}
      \nonumber \\ 
   &=& -2 \frac{(E+Q)(QE-\Delta^2)}{(E^2+\Delta^2)\left[(E+Q)^2 
      +\varepsilon (E^2+\Delta^2)\right]} \nonumber \\ 
   &=& -2 \frac{E}{E^2+\Delta^2}  \nonumber \\ 
   && \hspace{0cm} 
      + 2 \frac{E+\frac{Q}{1+\varepsilon}}{ 
      \left(E+\frac{Q}{1+\varepsilon}\right)^2 
      + \varepsilon 
      \left[ \frac{\Delta^2}{1+\varepsilon} 
      + \left(\frac{Q}{1+\varepsilon} \right)^2 \right]}.   
\EEQ{Phi}

\noindent Using the formula 

\BEQ 
   \frac{1}{\pi}
   \hat{\caP}\!\int_{-\infty}^{+\infty} \! dy  \;
      \frac{1}{y-x} \frac{y}{y^2+a^2} = \frac{|a|}{x^2+a^2}   
\EEQ{Hilbert}

\noindent for a real constant $a$, it follows 
from Eqs. (\ref{DensiConduFano2}) and (\ref{Phi}) that

\BEQ 
   \rho(E)   
   \simeq \frac{1}{\pi} \frac{|\Delta|}{(E-E_{0})^2+\Delta^2}.  
\EEQ{DensiConduFano} 

\noindent in $E\neq E_{0}-Q$. 


\section{Time reversal symmetry in the Fano resonance}  

   In this appendix we show $d_2=0$ under the conditions  
$A_{12}=A_{21}\neq 0$ and $S_{12}=S_{21}$.  
   This result $d_2=0$ leads to $\caK=0$, so we obtain $W_f=0$, 
$\Lambda_f=(q^2/(2\pi\hbar))|A_{12}|^2$ and $Q=- d_1$ by 
using Eqs. (\ref{ConduFano1}), (\ref{ConduFano2})
and (\ref{ConduFano3}).  

   The matrix $A$, which is an unitary matrix, 
is represented as 

\BEQ  
      A = 
      \left(
      \begin{array}{cc} 
         i e^{i(\tilde{\theta}+\tilde{\varphi}_1)} \sin{\tilde{\phi}} 
         & e^{i(\tilde{\theta}+\tilde{\varphi}_2)} \cos{\tilde{\phi}} 
         \\ e^{i(\tilde{\theta}-\tilde{\varphi}_2)} \cos{\tilde{\phi}} 
         & i e^{i(\tilde{\theta}-\tilde{\varphi}_1)} \sin{\tilde{\phi}}
      \end{array} 
      \right)
\EEQ{Smatr} 

\noindent with real parameters $\tilde{\theta}$, 
$\tilde{\varphi}_j, j=1,2$ and $\tilde{\phi}$.
   The condition $A_{12}=A_{21}$ imposes 
   
\BEQ 
  \tilde{\varphi}_2 = 0 \; \mbox{or} \; \pi.
\EEQ{TimeRevar1}  
   
\noindent The condition $S_{12}=S_{21}$ under Eq. (\ref{TimeRevar1}) 
implies that the multiplied matrix $AB$ is also symmetric. 
   This leads to the condition 
   
\BEQ 
   \tan{\phi} = 
   - \frac{\sin(\varphi+\tilde{\varphi}_1)}{\cos\tilde{\varphi}_2} 
   \tan\tilde{\phi}. 
\EEQ{TimeRevar2}   

\noindent On the other hand the constant $d_2$ given by Eq. (\ref{d2}) 
is represented as  

\BEQ 
   d_2 &=& - \Delta \cdot \cos{\phi} \cdot \Bigl[ 
      \tan{\phi}  \nonumber \\ 
    &&  \hspace{1cm} + \sin(\varphi+\tilde{\varphi}_1 
      - \tilde{\varphi}_2) \cdot \tan\tilde{\phi} \Bigr]. 
\EEQ{ConstD} 
   
\noindent By using Eqs. (\ref{TimeRevar1}), 
(\ref{TimeRevar2}) and (\ref{ConstD}) we obtain $d_2=0$.



\end{multicols}


\begin{thebibliography}{99}
\bibitem{note}   Strictly speaking from a mathematical point of view,  
the right-hand side of Eq. (\ref{DensiCondu2}) 
gives a delta functional singularity in the conductance zero 
point $E=E_{0}-Q$, but this singularity is not
a real contribution to the density of states. 
   (Here it should be emphasized again that the density of states 
is a continuous function of energy in the open conductor.) 
   Therefore we neglect this singularity in the 
results of calculation, and define the value of the 
density of states at the energy $E_{0}-Q$ as 
$\rho(E_{0}-Q)\equiv\lim_{E\rightarrow E_{0}-Q}\rho(E)$.
\bibitem{note2} Eq. (\ref{DensiState}) implies that the constant 
$\Delta$ must be positive because the density of states given 
by Eq. (\ref{DensiState}) must be positive. 
   Therefore the scattering matrix as a function 
of $E$ is analytic in the whole upper half 
plain and in the real axis in the imaginary number $E$.
\end{thebibliography}
\end{document}